\documentclass[useAMS,usenatbib]{mn2e}

\usepackage{graphics}
\usepackage{psfig}
\usepackage{amsmath}
\usepackage{amssymb}
\usepackage{upgreek}
\usepackage{longtable}
\usepackage[mathscr]{eucal}
\usepackage[dvips]{graphicx}
\usepackage{longtable}
\usepackage{lscape}
\usepackage{float}

\title
[Rotation of NGC 2516]
{3D Rotation of the Open Cluster NGC 2516}
\author
[Wright, Jeffries \& Whalley]
{Nicholas J. Wright$^1$, R.D. Jeffries$^1$, M.A. Whalley$^{1,2}$\\
\\
$^{1}$Astrophysics Research Centre, Keele University, Keele, ST5 5BG, UK\\
$^{2}$University of Chester, Chester, CH1 4BJ, UK\\
}

\begin{document}
\maketitle

\begin{abstract}

We have combined {\it Gaia} astrometry with {\it Gaia}-ESO Survey radial velocities to measure the 3D rotation of the open cluster NGC~2516. We compiled a sample of 376 members with astrometry and spectroscopy and use these to determine a distance to the cluster of $406.4 \pm 0.8$~pc, which we then use to infer the 3D positions and velocities of all stars in the cluster using a Bayesian model. We identify the axis of maximum cluster rotation and measure a median rotational velocity of $0.11 \pm 0.02$ km~s$^{-1}$. We find the axis of maximum cluster rotation to be $52^\circ \pm 22^\circ$ to the plane of our galaxy. We compare this rotation rate to measurements of cluster rotation in other open clusters and find that it is inconsistent with the expected dependences on cluster age and mass.

\end{abstract}


\section{Introduction}

Open clusters are gravitationally bound groups of hundreds to hundreds of thousands of stars with ages from several million to several billion years. They are thought to form from dense molecular clouds, exist briefly as young embedded clusters, and to navigate the residual gas expulsion and other disruptive processes responsible for unbinding young clusters, to then survive to older ages \citep{lada03,bast06,wrig20}. Open clusters are excellent astronomical laboratories for studying star and star cluster formation \citep{krou01}, stellar structure and evolution \citep[e.g.,][]{barn07,jack10}, and are good tracers for unveiling the structure and evolution of our Galaxy \citep{cast21,pogg21}. Despite this, the formation, evolution and eventual dispersal of open clusters is poorly understood.

The internal kinematics of open clusters have been sparsely studied, particularly relative to that of young clusters and associations \citep[e.g.,][]{wrig24}. This is primarily because the velocity dispersions and levels of systematic rotation in open clusters are small compared to the kinematic measurement uncertainties, particularly for radial velocities, and therefore the difficulty of obtaining high-precision radial velocities and proper motions for sufficient numbers of cluster members has hindered kinematic studies. As a result of this, very little is known about the rotation, expansion or contraction, level of equipartition or evolving dynamics of open clusters, with perhaps the exception of the nearest systems such as the Pleiades or the Hyades \citep[e.g.,][]{lodi19,evan22,hao24}. The internal kinematics of star clusters are initially set during their formation and so can provide constraints on how star clusters form \citep{pros09,wrig14b,wrig19b}, while their later evolution is driven by internal dynamics and Galactic tidal forces, thus providing constraints on the mass distribution within the Galaxy \citep[e.g.,][]{giel06}.

Hydrodynamical simulations show that young star clusters should inherit the angular momentum present in molecular clouds \citep[e.g.,][]{vesp14,lee16a,mape17}, leading to strong rotation that should be higher for clusters formed from hierarchical mergers and for more massive clusters \citep{mape17}. N-body simulations suggest that this rotation should be retained for a long time \citep{tion17}, though the magnitude of rotation should decrease due to the evaporation of the cluster \citep[e.g.,][]{erns07}.

Early evidence for open cluster rotation was found using Hipparcos data \citep[e.g.,][]{perr98,vere13}. Some studies have found evidence for rotation in open clusters using {\it Gaia} data \citep[e.g.,][]{kama19,hao22,guil23}, though the studies have been sparse and the data and methods adopted have differed between studies, making comparisons difficult. To accurately measure the 3D rotation of a cluster one needs both the 3D velocities of stars in the cluster (derived from proper motions and radial velocities), as well as their 3D positions. While {\it Gaia} does provide parallaxes, the distances derived from these measurements can have large uncertainties and so some degree of inference is required to properly derive the 3D structure of the cluster.

NGC 2516, sometimes known as the Southern Beehive Cluster, is an open cluster in the constellation of Carina \citep{jeff01}. It is located at a distance of $411 \pm 6$ pc \citep{jack22} and has an age either similar to or older than the Pleiades cluster \citep{frit20,dias21,jeff23}. In this work we adopt the lithium-based age of $138^{+7}_{-3}$ Myrs \citep{jeff23}. The tidal radius has been estimated as 13.5~pc \citep{hunt24}. Measurements of the mass of NGC~2516 range from $M = 1168.4 \pm 19.4$ M$_\odot$ \citep{alme25} and $\sim$1240--1560 M$_\odot$ \citep{jeff01} at the low end up to $M \sim 3500$ M$_\odot$ \citep{hunt24} at the high end. Despite this mass uncertainty, NGC~2516 is regularly found to be several times more massive than the similarly-aged Pleiades cluster with these three studies finding mass ratios of 1.8--3.7 between NGC 2516 and the Pleiades.

This study uses 3D positions and velocities derived from the combination of {\it Gaia} astrometry and radial velocities from the {\it Gaia}-ESO Survey to study the rotation of NGC~2516. In Section~2 we present the observational data used and the compilation of the sample of cluster members studied. In Sections~3 we measure the distance to the cluster and recreate its 3D structure with an inference-based approach. In Section~4 we identify and measure the rotation of the cluster and in Section~5 we discuss our findings.

\section{Observational Data}

\begin{figure*}
\centering
\includegraphics[width=400pt]{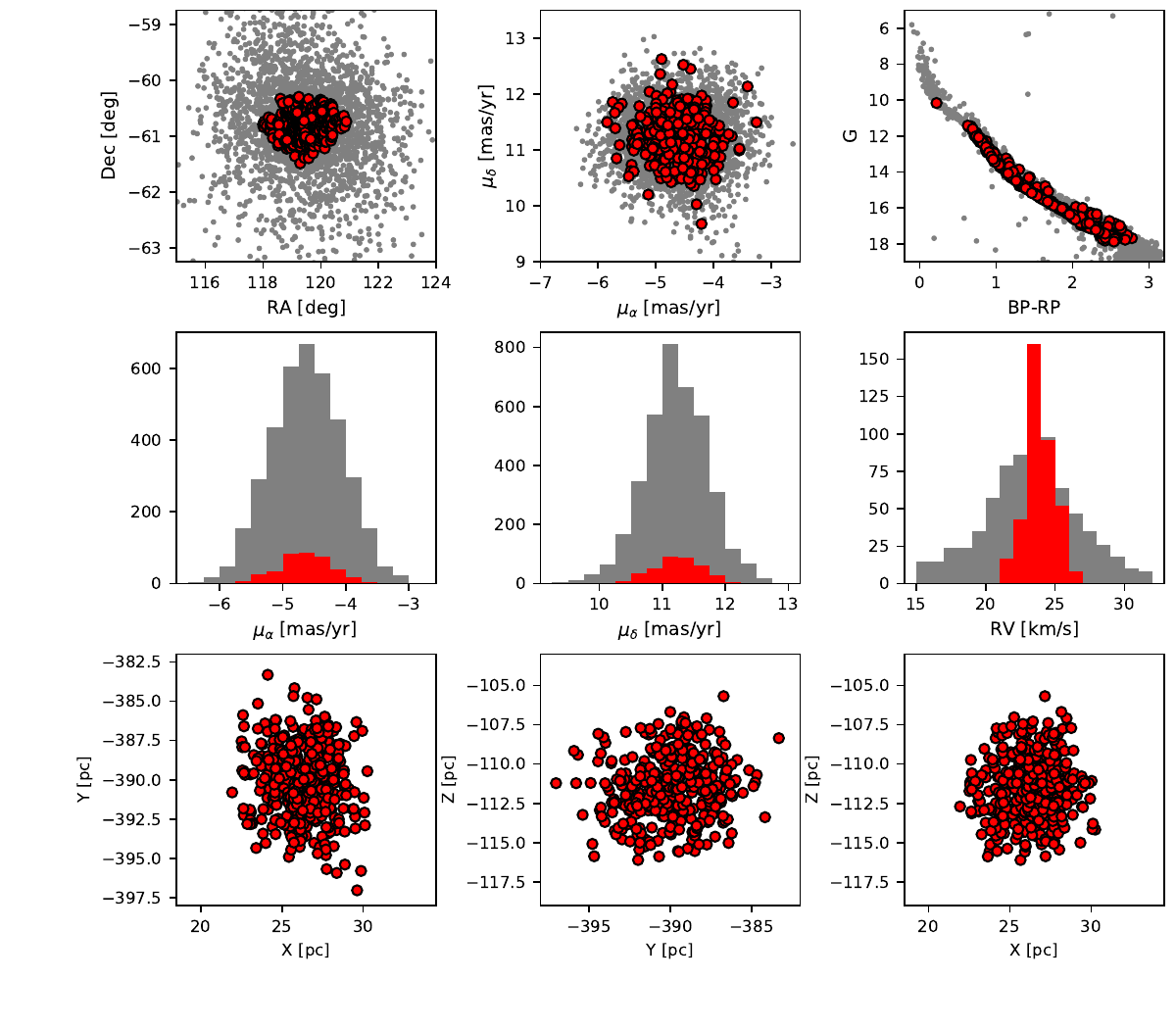}
\caption{Properties of our cluster members sample in NGC~2516 (red) compared to the members from \citet{hunt24}. The top row show the observed plane of the sky distribution (top left), proper motion distribution (top centre), and position in the {\it Gaia} colour-magnitude diagram (top right). The middle row shows histograms of the observed distribution of the kinematic measurements $\mu_\alpha$ (middle left), $\mu_\delta$ (middle centre) and RV (middle right). The bottom row shows the reprojected distribution of members in the 3D Galactic Cartesian coordinate system $XYZ$. The extent covered by the three $XYZ$ axes is the same in all three plots, 16~pc.}
\label{basic_info_plot}
\end{figure*}

The data used for this work is a combination of spectroscopy from the {\it Gaia}-ESO Survey and astrometry from {\it Gaia}. The {\it Gaia}-ESO Survey \citep[GES,][]{gilm22,rand22} observed 62 young and open clusters, as well as field stars in our galaxy, measuring radial velocities (RVs) and chemical abundances from high-resolution Very Large Telescope (VLT) Fibre Large Array Multi Element Spectrograph (FLAMES) spectroscopy. 759 stars were observed towards NGC~2516, of which 491 were classified as ``highly probable cluster members'' by \citet{jack22} with a 3D kinematic membership probability, $P > 0.9$ and have usable RVs. We identify and remove RV outliers, which are probably dominated by binary systems, by removing 3$\sigma$ RV outliers relative to the median RV of 23.82 km~s$^{-1}$, where $\sigma = 1.07$ km~s$^{-1}$ (calculated from the interquartile range). This leaves 376 stars, which constitutes the sample of cluster members used in this study\footnote{While it is not generally recommended to use a kinematically-defined membership selection method for a kinematic study of a cluster (as kinematic outliers in the cluster will not pass the membership selection and will be removed from the sample), this should not significantly affect our analysis since the kinematic outliers rejected by such a membership selection are unlikely to be bound to the cluster and therefore would not be useful for probing the cluster's rotation.} and critically provide the RVs used.

All 376 member stars have {\it Gaia} EDR3 astrometry \citep{brow20}, including positions, parallaxes, $\varpi$, and proper motions (PMs), $\mu_\alpha$ and $\mu_\delta$, that pass the requirement that the 're-normalised unit weighted error' be $\leq$ 1.25, commonly used as an astrometric goodness-of-fit indicator \citep{peno22}. The parallaxes were individually corrected for the non-zero {\it Gaia} parallax zero-point using the Python package \texttt{gaiadr3\_zeropoint} \citep{lind21}. Combining GES spectroscopy with {\it Gaia} astrometry provides 3D positions and velocities for all stars in our sample. The median parallax uncertainty of our sample is 0.03~mas, while the median PM and RV uncertainties are 0.039 mas yr$^{-1}$ ($\mu_\alpha$, equivalent to 0.075 km s$^{-1}$ at a distance of 406.4 pc), 0.037 mas yr$^{-1}$ ($\mu_\delta$, equivalent to 0.071 km s$^{-1}$) and 0.42 km s$^{-1}$ (RV), meaning that the cluster velocity dispersion is well-resolved in 3D \citep[0.7--0.9 km s$^{-1}$ in each dimension,][]{jack22}. Figure~\ref{basic_info_plot} shows the distribution of the cluster members in various diagrams.

\section{Reprojection and 3D Structure of the Cluster}

To study the 3D rotation of the cluster we first need to reproject the cluster into a 3D cartesian reference frame, and to do that we need to calculate the distance to the cluster and then reproject the individual parallaxes into individual distances.

\subsection{Distance to the cluster}

To calculate the distance to the cluster we fit a model to the observed parallax distribution of the 376 member stars using a forward model. We assume that all of the member stars are distributed in a single cluster with a line-of-sight distance distribution that can be approximated by a gaussian with central distance, $d_{cluster}$, and a dispersion, $\sigma_{cluster}$. The forward model calculates a distribution of distances that are reprojected into parallax space and to which the effects of parallax uncertainty are added by randomly sampling from a second gaussian with centre equal to the modelled parallax and a standard deviation sampled from the observed parallax uncertainties of the 376 member stars.

This forward model produces a modelled parallax distribution that is compared to the observed parallax distribution using an unbinned maximum likelihood test, with likelihood function

\begin{equation}
\mathcal{L}(\theta \mid x) = \prod_{i=1}^{i = n} P (x_i)
\end{equation}

\noindent where $\theta = (d_{cluster}, \sigma_{cluster})$ are the model parameters, $x = \varpi$ are the observable parallaxes, and $P(x_i)$ is the probability density function described above and evaluated at the $n = 376$ observed data points. We set priors of $d_{cluster} = U(0,1000)$~pc and $\sigma_{cluster} = U(0, 10)$~pc.

We use the Markov Chain Monte Carlo (MCMC) code \texttt{emcee} to explore the posterior distribution. We used 1000 walkers and ran the code for 2000 iterations, with the first 1000 discarded as a burn-in, using the \texttt{emcee} autocorrelation time to ensure convergence. The median of the posterior distribution is taken as the best-fit, and the 16th and 84th percentiles used to calculate the 1$\sigma$ confidence intervals. We fit a distance of $406.4 \pm 0.8$~pc with a line of sight distance dispersion of $4.7 \pm 1.0$~pc. This distance is in reasonable agreement with the {\it Gaia} EDR3 distance of $411.1 \pm 6$ pc calculated by \citet{jack22}, while the dispersion is in good agreement with the plane-of-the-sky size of the cluster (as characterised from half of the 16$^{th}$ to 84$^{th}$ percentile range) of $\sigma_\alpha = 3.8$~pc and $\sigma_\delta = 4.1$~pc, calculated using the sample of cluster members from \citet{hunt24}.



\subsection{Reprojection of the cluster members}

To reproject the cluster into a 3D cartesian reference frame we first derive individual distances for all stars using an inference-based process with a prior that requires that all stars be part of a cluster at the distance derived earlier. We use a gaussian prior with centre of 406.4~pc and a dispersion of 4.7~pc, the parameters fitted in the previous section. This is used to constrain the distance to individual stars, which are forward modelled to produce a parallax that is compared to the observed parallax with a maximum likelihood test. The results of this process are distances for all stars that are consistent with them being part of a clustered distribution. These distances are then combined with the positions, proper motions and radial velocities to calculate 3D cartesian positions and velocities in the Galactic cartesian coordinate system using the Python AstroPy module \texttt{SkyCoord}.

Figure~\ref{basic_info_plot} shows the reprojected 3D distribution of our 3D cluster members in $XYZ$ space. As expected, the structure of our sample of members is elongated along the $Y$ axis, the axis closest to the line of sight. This is not necessarily due to the elongation of the apparent structure of the cluster by parallax uncertainties, but because the GES observations sample only the core of the cluster on the plane of the sky (see Figure~\ref{basic_info_plot}) but should sample the full extent of the cluster along the line of sight. This should not impact our analysis, but it is important to be aware of.

\section{Rotation of NGC 2516}

\begin{figure*}
\centering
\includegraphics[width=500pt]{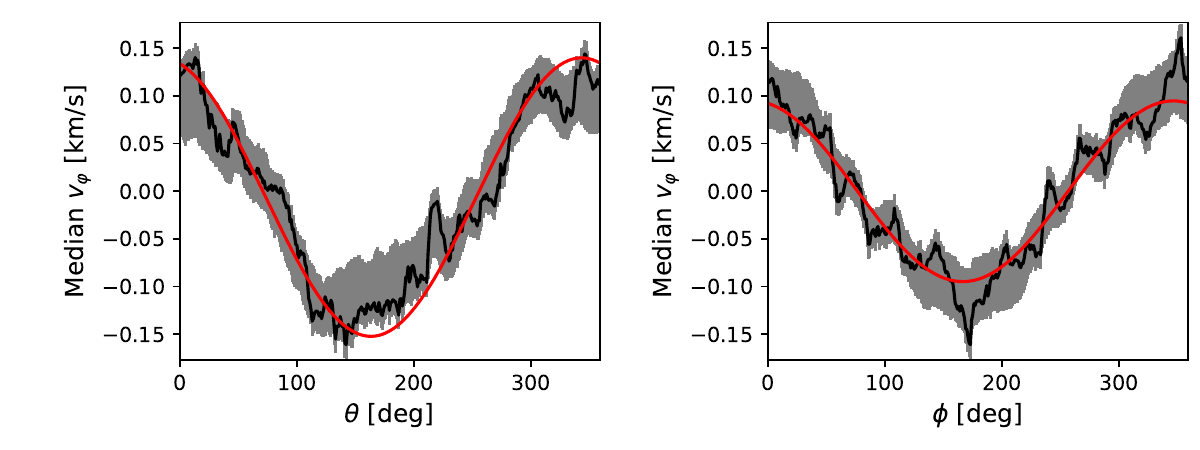}
\caption{Angular dependence of the median rotational velocity, $v_\varphi$, for a rotation axis orientated by the angles ($\theta, \phi$) relative to the Galactic cartesian coordinate system. The black lines show the median value of $v_\varphi$ at each angle, with the grey shaded area showing the uncertainty. In the left-hand panel, where $\theta$ is allowed to vary, $\phi$ is set to the value of maximum rotation of $\phi = 168^\circ$. In the right-hand panel, where $\phi$ is allowed to vary, $\theta$ is set to the value of maximum rotation of $\theta = 164^\circ$. The red line shows the fitted sinusoidal functions.}
\label{median_vperp}
\end{figure*}

To measure the rotation of NGC~2516 we must identify the axis around which the rotation of the cluster is maximised. Traditionally this has been performed using the residual velocity method that has been used previously by many authors \citep[e.g.,][]{bell12,lanz18,hao22}. The residual velocity method has most commonly been applied to globular clusters using data consisting of 2D positions (plane of the sky positions) and 1D velocities (RVs). The method involves finding the axis, in the plane of the sky, that maximises the velocity difference either side of that axis, making that axis the projection of a 3D rotation axis onto the plane of the sky (and therefore making the measured rotational velocity amplitude a lower limit on the true amplitude). With the availability of 3D positions and velocities, this method has been adapted to work in 3D \citep[e.g.,][]{hao22}, but it is not an ideal method as it only considers one dimension of position and velocity at a time. Our approach here is to identify the rotation axis of the cluster as that which maximises the 2D rotational velocities in the plane perpendicular to this axis.

\subsection{Methods}

With the 3D positions and velocities of all stars defined in the Galactic cartesian coordinate system, our first step is to reproject all stars into a similar cartesian coordinate system, on the same axes, but for which the centre of the cluster, in both positional and velocity space, defines the origin of the system. We do this by subtracting the median position or velocity from each respective axis.

To determine the rotation axis of the cluster, we  rotate our coordinate system by the angles $\theta$ and $\phi$, where we define $\theta$ as the rotation of the coordinate system in the $X-Z$ plane from the positive $Z$ axis towards the positive $X$ axis, and $\phi$ as the rotation of the coordinate system in the $X-Y$ plane from the positive $Y$ axis towards the positive $X$ axis. This produces a new coordinate system that we denote $X'Y'Z'$. At each combination of $(\theta, \phi)$ we reproject the positions and velocities into a cylindrical coordinate system with positions $(r, \varphi, z)$ and velocities $(v_r, v_\varphi, v_z)$, with the system aligned such that the polar axis, $z$, is aligned with the $X'$ axis of the new cartesian system. In this cylindrical system, $v_\varphi$ represents the rotational velocity of each star, i.e., the velocity component of the star in a plane perpendicular to the rotation axis and perpendicular to a radial line from the rotation axis to the star.

We explore all values of $\theta$ and $\phi$ from 0$^\circ$ to 360$^\circ$ in steps of 1$^\circ$, and determine the median value of $v_\varphi$ for each combination of $(\theta, \phi)$. The $\theta, \phi$ values where the median $v_\varphi$ is maximised are taken as a first estimate of the orientation of the axis of maximum rotation. A step size of 1$^\circ$ was found to be sufficient to identify the axis of maximum rotation and decreasing this did not change the result.

Due to the noise within the data we improve our estimate of the axis of maximum rotation by fitting a sinusoid to the dependence of the median value of $v_\varphi$ on both angles. We determine the dependence of the median of $v_\varphi$ on the two angles $\theta$ and $\phi$ by varying one angle over 360$^\circ$ while the other angle is kept constant as the angle of maximum rotation, giving the quantities $v_\varphi(\theta)$ and $v_\varphi(\phi)$ (see Figure~\ref{median_vperp}). We fit a sine wave to these quantities using the Python {\sc statsmodels} module and Ordinary Least Squares (OLS) regression. We use the peak angle of this sine wave for both $\theta$ and $\phi$ to define the rotation axis, and the amplitude of the sine wave as the rotational velocity of the cluster, a process that overcomes the noise within the data.

\subsection{Results}

For NGC~2516 we find the axis of maximum rotation to be at ($\theta, \phi) = (164^\circ, 168^\circ)$ with an uncertainty of $\pm22^\circ$, with a median rotational velocity of $0.11 \pm 0.02$ km~s$^{-1}$. Figure~\ref{median_vperp} shows the angular dependence of the median rotational velocity, where the sinusoidal behaviour provides a clear indication of the rotation of the cluster. The axis of maximum rotation is found at an angle where the median rotational velocity is negative, indicating anti-clockwise motion when looking down on the rotational plane. The angle between the rotation axis and the plane of our galaxy is $52^\circ \pm 22^\circ$ and the angle between the rotation axis and a line between the cluster and the Galactic Centre in the $X-Y$ plane is $54\circ \pm 29^\circ$. All uncertainties were calculated using Monte Carlo simulations, varying the observed values by their uncertainties, repeating the analysis and using the 16$^{th}$ and 84$^{th}$ percentiles for the uncertainties on all results.

\begin{figure}
\centering
\includegraphics[width=260pt]{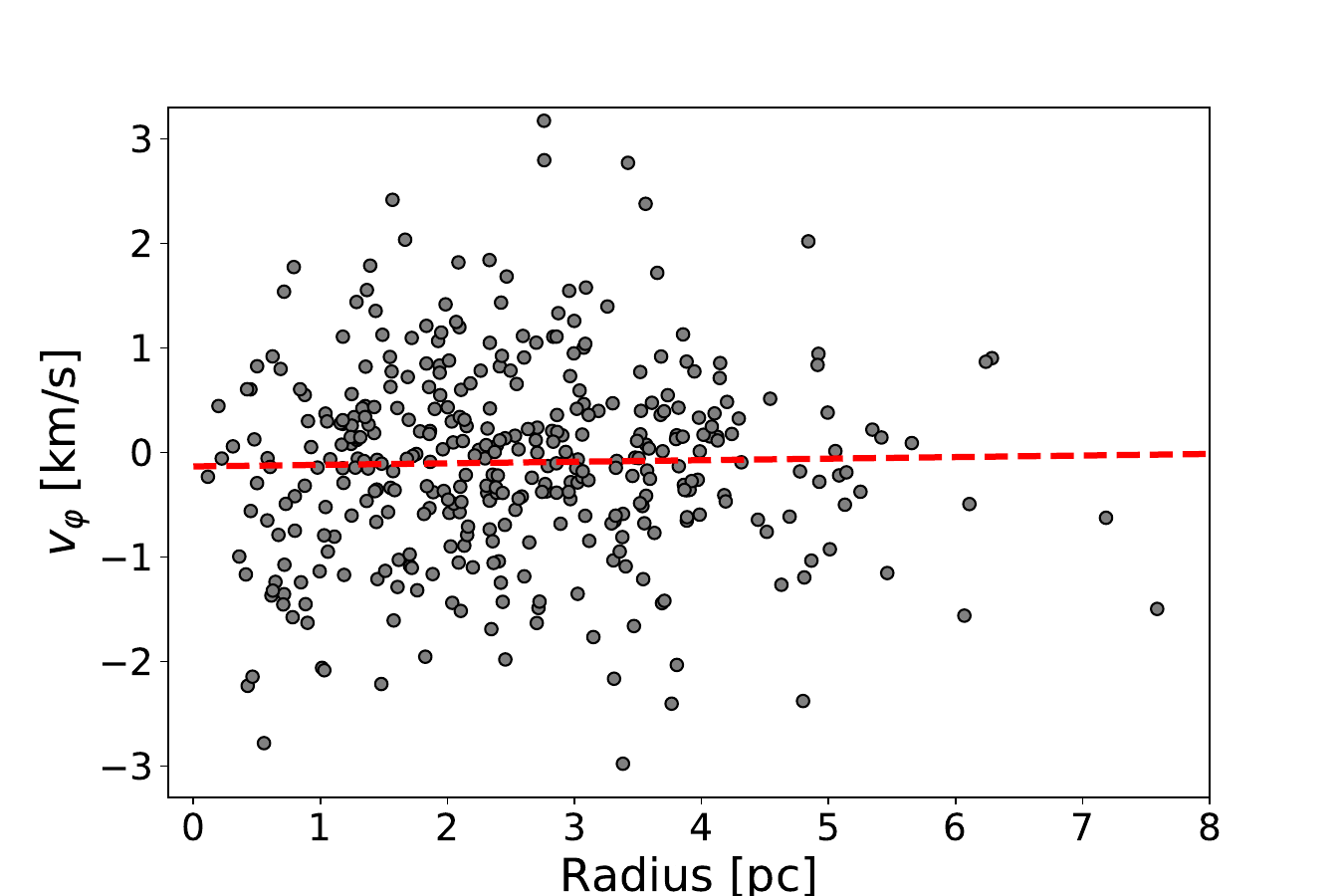}
\caption{Rotational velocity, $v_\varphi$, plotted against radius (in the plane perpendicular to the rotation axis) for all stars in our sample. The red dashed line shows the best-fitting linear relationship with a gradient of $0.015 \pm 0.036$ km~s$^{-1}$~pc$^{-1}$.}
\label{vperp_radius}
\end{figure}

Figure~\ref{vperp_radius} shows the rotational velocities, $v_\varphi$, of the cluster members as a function of their radii (the distance from the rotational axis). All of these stars are within the tidal radius of the cluster \citep[13.5~pc,][]{hunt24}. The distribution in $v_\varphi$ is much larger than the measurement uncertainties, indicating that the dispersion observed is real. To determine if there is any radial dependence on the rotational velocity we fit a linear function using Python's {\sc statsmodel} module and find a gradient of $0.015 \pm 0.036$ km~s$^{-1}$~pc$^{-1}$. This is consistent with no variation with radius, at least within the range of radii studied.


\section{Discussion}

We have measured for the first time the 3D rotation rate of the massive open cluster NGC~2516. We have measured a rotational velocity of $0.11 \pm 0.02$ km~s$^{-1}$ at an angle of $52^\circ \pm 22^\circ$ to the plane of our galaxy. We do not find any radial dependence on the rotational velocity with the range of radii studied, though our sample sits significantly within the tidal radius so some variation may be present a greater radii.

NGC~2516 joins a small group of open clusters with measured 3D rotation rates, which includes Alpha Persei, the Pleiades, Praesepe and the Hyades. The rotational velocity we have measured here is very similar to that of Praesepe \citep[$0.12 \pm 0.05$ km~s$^{-1}$,][]{hao24}, despite that cluster being much older \citep[$\sim$705 Myr,][]{lodi19a}. The Pleiades cluster is a similar age to NGC~2516, yet \citet{hao24} measured a rotational velocity of $0.24 \pm 0.04$ km~s$^{-1}$ for the Pleiades, which is much less massive than NGC~2516. While there is some uncertainty on the total mass of NGC~2516 \citep[e.g.,][]{jeff01,hunt24,alme25} it is definitely more massive than the Pleiades cluster, with the referenced studies finding mass ratios between the two clusters of 1.8--3.7, despite predictions that the magnitude of cluster rotation should increase with cluster mass \citep{vesp14,mape17}. In fact, NGC~2516 is much more massive than any of the clusters whose rotation rate has previously been measured with both \citet{hunt24} and \citet{alme25} finding it approximately 5--15 times more massive than the Hyades, 2-3 times more massive than Praesepe and 2--8 times more massive than Alpha Per. Despite this, many of these clusters have higher measured rotation rates than NGC~2516, for example \citet{hao24} measure a rotational velocity of $0.43 \pm 0.08$ km s$^{-1}$ for the similarly-aged Alpha Per cluster.

These differences may be due to the different methods used to identify and measure rotation between the two studies, the different radial extents of the different samples, or be due to their formation and evolutionary histories. For example the \citet{hao24} studies extend beyond the tidal radii in the clusters they studied, while our study is focussed on the core of NGC~2516, well within the tidal radius. If the rotational velocity does have a radial dependence, this could go some way to explaining this discrepancy. Further studies of the rotation of young and open clusters using homogeneous samples and methods will be needed to fully determine the dependence of cluster rotation on age and mass.

\section{Conclusions}

In this paper we have identified and measured the 3D rotation of the open cluster NGC~2516. We have used high-precision astrometry from {\it Gaia} EDR3 and RVs from the {\it Gaia}-ESO Survey for a sample of 376 members of the cluster. We have determined the distance to the cluster to be $406.4 \pm 0.8$~pc and reprojected the positions and parallaxes of the members using a Bayesian model to reconstruct the 3D structure of cluster. Using our 3D velocities we have identified the axis of maximum cluster rotation and measured rotational velocities, $v_\varphi$, perpendicular to this axis for all members. We find the axis of maximum cluster rotation to be $52^\circ \pm 22^\circ$ to the plane of our galaxy and the median rotational velocity to be $0.11 \pm 0.02$ km~s$^{-1}$. This is notably lower than that recently measured for the similarly-aged Pleiades cluster, suggesting that factors other than age may be important for determining the rotation rate of a cluster. Nonetheless, this study highlights the potential for detailed studies of the internal kinematics of open clusters possible with the combination of {\it Gaia} astrometry and RVs from multi-object spectroscopic surveys.

\section*{Acknowledgments}

This research has made use of NASA's Astrophysics Data System and the Simbad and VizieR databases, operated at CDS, Strasbourg. This work is based on data products from observations made with ESO Telescopes at the La Silla Paranal Observatory under program ID 188.B-3002. These data products have been processed by the Cambridge Astronomy Survey Unit (CASU) at the Institute of Astronomy, University of Cambridge, and by the FLAMES-UVES reduction team at INAF-Osservatorio Astrofisico di Arcetri. These data have been obtained from the Gaia-ESO Survey Data Archive, prepared and hosted by the Wide Field Astronomy Unit, Institute for Astronomy, University of Edinburgh, which is funded by the UK Science and Technology Facilities Council. This work also made use of results from the European Space Agency (ESA) space mission Gaia. Gaia data are being processed by the Gaia DPAC. Funding for the DPAC is provided by national institutions, in particular the institutions participating in the Gaia MultiLateral Agreement (MLA). The Gaia mission website is https://www.cosmos.esa.int/gaia. The Gaia archive website is https://archives.esac.esa.int/gaia.

\section*{Data Availability}

The GES and {\it Gaia} data used in this article are available from the ESO and Gaia archives, respectively. The data underlying this article will be made available in the Vizier archive upon publication of this article.

\bibliographystyle{mn2e}
\bibliography{/Users/nwright/Documents/Work/tex_papers/bibliography.bib}
\bsp

\end{document}